\documentclass[reprint,superscriptaddress,amsmath,amssymb,amsthm,amsfonts,aps,pra]{revtex4-2}

\usepackage{graphicx}
\usepackage{dcolumn}
\usepackage{bm}
\usepackage[mathlines]{lineno}

\usepackage{braket}
\usepackage{physics}
\usepackage{mathtools}
\usepackage{tikz-cd}
\usepackage{dsfont}
\usepackage{scalerel}
\usepackage{ulem} 

\usepackage{mathbbol}
\DeclareSymbolFontAlphabet{\amsmathbb}{AMSb}%
\newtheorem{theorem}{Theorem}
\newtheorem{lemma}{Lemma}

\newtheorem{corollary}{Corollary}

\def\Rset{\mathds{R}}

\newcommand{\tp}{\intercal}
\def\ie{\textit{i.e., }}

\def\Bell{\mathrm{Bell}}
\def\ma{\mathbb{a}}
\def\ba{\bar{a}}
\def\a{a}
\def\b{b}

\def\mLambda{\mathbb{\Lambda}}

\def\diag{\text{diag}}
\def\mid{\mathbb{1}}
\def\mr{\mathbb{r}}
\def\mrd{\mathbb{r}_{\rm d}}
\def\mrtd{\tilde{\mathbb{r}}_{\rm d}}
\def\mo{\mathbb{o}}
\def\mb{\mathbb{b}}
\def\mw{\mathbb{w}}
\def\m0{\mathbb{0}}
\def\mR{\mathbb{R}}
\def\hsigma{\hat\sigma}
\def\hbsigma{\hat{\boldsymbol{\sigma}}}
\def\homega{\hat{\omega}}

\def\hrho{\hat{\rho}}
\def\hbeta{\hat{\beta}}
\def\hrhoc{\hat{\rho}_{\rm c}}
\def\hid{\hat{\mathbb{1}}}
\def\inp{\mathrm{in}}
\def\out{\mathrm{out}}
\def\mC{\mathbb{C}_{\scriptscriptstyle{\Lambda}}}
\def\mCal{\mathbb{C}_{\scriptscriptstyle{\Lambda^{\rm alig}}}}

\def\mwmd{\mathbb{w}_{{\rm d}m}}

\def\mwtmd{(\tilde{\mathbb{w}}_m)_{{\rm d}}}
\def\vn{{\bf n}}
\def\vt{{\bf t}}
\def\va{{\bf a}}
\def\vr{{\bf r}}
\def\vrc{{\bf r}_{\rm c}}
\def\vzero{\mathbf{0}}
\def\vv{{\bf v}_{\scriptscriptstyle\Lambda}}
\def\vvv{{\bf v}}
\def\vka{\boldsymbol{\kappa}}
\def\vw{\boldsymbol{\omega}}

\def\vv{{\bf v}_{\scriptscriptstyle\Lambda}}
\def\B{\mathcal{B}}

\newcommand{\beq}{\begin{equation}}
	\newcommand{\eeq}{\end{equation}}
\newcommand{\bse}{\begin{subequations}}
	\newcommand{\ese}{\end{subequations}}
\newcommand{\bea}{\begin{eqnarray}}
	\newcommand{\eea}{\end{eqnarray}}

\usepackage{xcolor}
\usepackage{soul}

\usepackage{changes}
\definechangesauthor[name=Diego,color=green]{diego}
\definechangesauthor[name=Ana,color=brown]{ana}

\begin{document}

\title{Optimal quantum teleportation protocols for fixed average fidelity}

\author{Fabricio Toscano}  
\email{toscano@if.ufrj.br}
\affiliation{Instituto de F\'isica, Universidade Federal do Rio de Janeiro, 21941-972, Rio de Janeiro, Brazil}

\author{Diego G. Bussandri}
\email{diego.bussandri@unc.edu.ar}
\affiliation{Instituto de F\'isica La Plata (IFLP), CONICET--UNLP, 1900 La Plata, Argentina}

\author{Gustavo M. Bosyk}
\email{gbosyk@fisica.unlp.edu.ar}
\affiliation{Instituto de F\'isica La Plata (IFLP), CONICET--UNLP, 1900 La Plata, Argentina}

\author{Ana P. Majtey}
\email{anamajtey@unc.edu.ar}
\affiliation{Instituto de Física Enrique Gaviola (IFEG), CONICET--UNC, C\'ordoba, Argentina}
\affiliation{Facultad de Matemática, Astronomía, Física y Computación, Universidad Nacional de Córdoba,  
X5000HUA Córdoba, Argentina}

\author{Mariela Portesi}
\email{portesi@fisica.unlp.edu.ar}
\affiliation{Instituto de F\'isica La Plata (IFLP), CONICET--UNLP, 1900 La Plata, Argentina} 
\affiliation{Facultad de Ciencias Exactas, Universidad Nacional de La Plata, 1900 La Plata, Argentina}

\date{\today}

\begin{abstract}
We demonstrate that among all quantum teleportation protocols giving rise to the same average fidelity, those with aligned Bloch vectors between input and output states exhibit the minimum average trace distance. This defines optimal protocols. Furthermore, we show that optimal protocols can be interpreted as the perfect quantum teleportation protocol under the action of correlated one-qubit channels. In particular, 
we focus on the deterministic case, 
for which the final Bloch vector length is equal for all measurement outcomes. Within these protocols, there exists one type that corresponds to the action of uncorrelated channels: these are depolarizing channels. Thus, we established the optimal 
quantum teleportation protocol under a very common experimental noise.
\end{abstract}

\maketitle

\section{Introduction} 

Among the most astonishing techniques in quantum information theory are the  quantum teleportation protocols (QTPs), which consist of two distant parties, usually called Alice and Bob, aiming to transmit an unknown qubit state $\hrho^{\inp}$ from Alice qubit system $\ba$ to Bob qubit system $\b$, exploiting the features of quantum states and quantum measurements \cite{Bennett1993}.

QTPs are a paradigmatic example of local operations and classical communication (LOCC) protocols, defined on a system composed of three qubits: the system $\ba$, an additional qubit $a$, and the target system $b$ \cite{Popescu1994}. 
The most general teleportation protocol operates on the total system $\hrho^{\inp}\otimes \hrho^{ab}$, where the joint state $\hrho^{ab}$ is usually referred to as the resource state. 
The protocol goes as follows: first Alice performs a joint measurement on her qubits $\ba$ and $\a$, followed by the classical communication of the corresponding measurement outcome (labelled by $m$) to Bob, who finally applies local unitary operations on his qubit $\b$ according to the communicated result. %
The noiseless standard quantum teleportation is the only scheme that allows perfect transmission, {\it i.e.}  $\hrho^{\out}_m=\hrho^{\inp}$ $\forall\, m$ and for any input state, being $\hrho^{\out}_m$ the output states in the target system $\b$ \cite{Bennett1993,Horodecki1996bis,Horodecki2001book}. This protocol consists of a Bell measurement, \ie a projection onto the Bell basis $\{\hat{\beta}_m\}_1^4$ on qubits $\ba$ and $\a$, and a Bell state as quantum resource, $\hrho^{ab}=\hbeta$.   

In realistic teleportation implementations, states and measurements are typically not perfect.
The average fidelity between input and output states is generally employed as a figure of merit of the transmission process \cite{Horodecki1996bis,Vidal2000,Horodecki2001book,Albeverio2002}. 
In noisy standard QTPs, Alice implements a  Bell  measurement, where the resource state $\hrho^{\a\b}$ is taken to be an arbitrary mixed state.  
Within these standard protocols, one approach is to maximize the average fidelity over all Bob unitary operations, the so-called strategies, to determine what kind of mixed resource states give rise to quantum teleportation, {\it i.e.} when the average fidelity exceeds the bound $\frac{2}{3}$ for classical teleportation  \cite{Horodecki1996bis,Gisin1996}. 
Another approach is, for any given initial resource state, to maximize the singlet fraction, {\it i.e.} the fidelity between the resource state and the singlet Bell state, by LOCC, in order to produce a state $\hrho^{\a\b}$ with the highest average fidelity, to be used with  the standard QTP \cite{Horodecki1999,Horodecki2001book}. These are called optimal standard QTPs. 

Furthermore, for general resource states and positive operator-valued measures (POVMs), the optimal protocol was given in Ref.~\cite{Taketani2012} using the same framework for the average fidelity as in Ref.~\cite{Horodecki1999}.
However, as we show below, we identify several protocols that give rise to the same average fidelity, but that can produce significantly different output states. 
In Ref.~\cite{Bina2014} the limited effectiveness of fidelity as a tool for evaluating quantum resources was demonstrated. 
Here, we employ the trace distance as an additional quantum distinguishability measure to define the set of optimal QTPs, 
in the following sense: they  minimize the average trace distance for a fixed value of the average fidelity. 
One of our main findings is showing that this set is given by the teleportation protocols that align, \ie those for which the direction of the Bloch vector of the output states is the same as that of the initial state to be teleported. 

\section{General teleportation protocols.} 

Let us introduce the main elements for our analysis and fix the notation.
The input state of Alice  qubit system $\ba$ can be written as $\hrho^{\inp} =\frac{1}{2}\left(\hid+\vt^\tp\hbsigma\right)$, 
where $\vt=(t_1,t_2,t_3)^\tp$ is the Bloch vector of $\hrho^{\inp}$ with euclidean norm $t=\|\vt\|\leq 1$, $\hbsigma=(\hsigma_1,\hsigma_2,\hsigma_3)^\tp$ is the vector of Pauli operators, $\cdot^\tp$ denotes transposition, and $\hid$ is the identity operator. 
The resource state can be written as 
\beq
\hrho^{ab}\!=\!\frac{1}{4}\!\left(\hid_4 +(\vr^{\a})^\tp \hbsigma \otimes\hid\!+\!\hid \otimes (\vr^{b})^\tp \hbsigma\!+\!\sum_{i,j=1}^3\!\mr_{ij}\hsigma_{i}\otimes\hsigma_{j}\right)
\label{eq:Fano2}
\eeq
where $\vr^{\a}$ and $\vr^{b}$ are, respectively, the  Bloch vectors of the reduced states $\hrho^{\a} = \Tr_{\b}( \hrho^{\a\b})$ and $\hrho^{\b} = \Tr_{\a} (\hrho^{\a\b})$, and $\mr_{ij}$ are the elements of the correlation matrix $\mr = \Tr\left(\hrho^{ab} \,\hbsigma \otimes \hbsigma \right)$. The parametrization \eqref{eq:Fano2} defines the Fano form of a two-qubit state \cite{Fano1983}.
\par
We shall consider general measurements on Alice qubit systems $\ba$ and $\a$ described by  POVMs, that is, a set $\{\hat{E}_m^{\ba \a}\}$ of positive-definite operators  acting on the Hilbert space ${\cal H}^{\ba \a}$ such that $\sum_{m} \hat{E}^{\ba\a}_m=\hid \otimes \hid$. 
Each POVM element $\hat{E}_m^{\ba\a}$  defines univocally a two-qubit POVM state by means of $\homega^{\ba\a}_{m} = \frac{1}{4\bar{P}_m}\,\hat{E}^{\ba\a}_m$ where  $\bar{P}_m=\frac{1}{4}\Tr(\hat E_m^{\ba\a})$.
Each POVM state $\homega^{\ba\a}_{m}$ is completely characterized by its Fano form,  in terms of the Bloch vectors $\vw^{\ba}_m$ and $\vw^{\a}_m$ of the reduced states $\homega_m^{\ba} = \Tr_{\a}( \homega_m^{\ba\a})$ and 
$\homega_m^{\a} = \Tr_{\ba}( \homega_m^{\ba\a})$, respectively, and the correlation matrix 
$\mw_m= \Tr(\homega_m^{\ba\a} \,\hbsigma \otimes \hbsigma)$.
Note that because the POVM elements add up to the identity, the following POVM conditions have to be fulfilled: 
\bse
\label{eq:POVMcon}
\bea
&&\sum_{m}\,\bar{P}_m=1 , 
\label{eq:POVMcona}\\
&&\sum_{m} \bar{P}_m (\vw^{\a}_m)^\tp={\bf 0}^\tp,
\label{eq:POVMconb}\\ 
&&\sum_{m} \bar{P}_m\vw^{\ba}_m ={\bf 0}, 
\label{eq:POVMconc}\\
&&\sum_{m} \bar{P}_m \mw_m =\m0,
\label{eq:POVMcond}
\eea
\ese
where $\bf 0$ and $\m0$ denote the null vector and null matrix, respectively.  
\par
As a result of Alice measurement, the qubit $\b$ of Bob collapses to 
$\hrho^{b}_m = \frac{1}{2}\left[\hid+ (\vt^{b}_{m})^\tp\hbsigma \right]$ with probability
$P_m =  \Tr(\hat E_m^{\ba\a}\otimes\hid^\b
\hrho^{\inp}\otimes \hrho^{\a\b})=\bar{P}_m \,g_m(\vt)$, where  
$g_m(\vt) =1+(\vw_m^{\a})^\tp\vr^{\a}+ (\mw_m \vr^{\a}+\vw_m^{\ba})^\tp\vt$.
The Bloch vector of $\hrho^{b}_m$  is  
\beq
\label{eq:tmb}
    \vt^b_{m}= \frac{\ma_m}{g_m(\vt)} \,\vt + \frac{\vka_m}{g_m(\vt)} , 
\eeq 
where $\ma_m=\vr^b(\vw^{\ba}_m)^\tp+\mr^\tp \mw_m^\tp$ \ and \ $\vka_m=\vr^b+\mr^\tp\vw_m^a$. 
Finally Alice communicates to Bob her measurement result, $m$, and Bob applies a unitary operation, $\hat U_m$, on qubit~$b$. The output quantum state is \ $\hrho^{\out}_m  =\hat{U}_m\hrho^{b}_{m}\hat U_m^\dagger=\frac{1}{2}\left(\hid+\vt_m^\tp\hbsigma\right)$ with Bloch vector 
\beq
\label{eq:tm}
\vt_m= \mR_m \vt^b_{m},
\eeq
where $\mR_m$ is the unique rotation matrix such that $\hat{U}_m\, \vn^\tp\hbsigma \, \hat{U}_m^\dagger = 
\left(\mR_m \vn \right)^\tp \hbsigma$ with $\vn$ a unit real column vector. 
Thus, for each QTP there is an associated channel $\Lambda$ that yields  
$\Lambda(\hrho^{\inp})=\sum_m P_m \hrho^{\out}_m$ whose Bloch vector is 
\begin{equation*}
\vt_{\Lambda}=\sum_m P_m \vt_m=\mC\vt+\vv, 
\end{equation*}
with $\mC=\sum_m\bar{P}_m\mR_m\ma_m$ \ and \  $\vv=\sum_m \bar{P}_m \mR_m \vka_m$.
\par

\hfill

\section{Generalized error measures in quantum teleportation.}
 
The performance of a general QTP can be quantified by taking a measure of distinguishability between the input state and the ensemble of output states,   in the form  
$\bar{D}(\hrho^\inp) = \sum_{m} P_m \,D(\hrho^\inp, \hrho^\out_m)$ where  $D(\cdot,\cdot)$ stands for a distance measure between quantum states. Being $P_m=\bar{P}_m \, g_m(\vt)$, for any choice of $D$ the previous quantity can be expressed as a function of the initial Bloch vector $\vt$, so we write $\bar{D}(\hrho^\inp)=\bar{D}(\vt)\equiv \bar{D}$.
\par
The final figure of merit is the average distance defined as the expectation value of $\bar{D}$ over the uniform distribution of pure input states:
$\expval{\bar{D}}
=\frac{1}{4\pi} \iint_{S(\B)} \, \bar{D}(\vt) \, d\Omega$, where $d\Omega=\sin\theta\,d\theta\,d\phi$ \ ($0\leq\theta\leq \pi $ and $0\leq \phi < 2\pi$) is the differential solid angle in the Bloch sphere $S(\B)$.
The distance deviation $\Delta \bar{D}$ is defined as the standard deviation of the function $\bar{D}$, that is,
$\Delta \bar{D} =\sqrt{\expval{\bar{D}^2 \rangle-\langle \bar{D} }^2}$.
\par
In this work, we shall consider the following  distance measures: 
the trace distance  $D_{\text{T}}(\hrho, \hsigma)=\frac{1}{2}\Tr(\|\hrho-\hsigma\|)$ where $\|\hat A\|=\sqrt{\hat A\hat A^\dagger}$ stands for the operator norm \cite{NielsenChuangBook}, 
and 
the Uhlmann--Jozsa quantum fidelity 
$F(\hrho, \hsigma)= \left(\Tr ( \sqrt{\sqrt{\hrho} \, \hsigma \sqrt{\hrho}} )  \right)^2 $ 
\footnote{We notice that, strictly speaking, fidelity itself is not a metric on density operators but gives rise to one, see e.g.~\cite{NielsenChuangBook}.}.
For qubit states characterized by Bloch vectors $\vt$ and $\vt_m$, they give \ 
$D_{\text{T}}(\hrho^\inp, \hrho^\out_m)=\frac{1}{2}\|\vt-\vt_m\|$, and 
 $F(\hrho^\inp, \hrho^\out_m)=\frac{1}{2}
(1+\vt^\tp \vt_m+\sqrt{1-t^2}\sqrt{1-t^2_m})$  
where $t=\|\vt\|$ and $t_m=\|\vt_m\|$.
\par
The average fidelity takes the following form for general QTPs: 
\beq
\expval{\bar{F}}=\frac{1}{2}\left[1+\frac{1}{3}\tr(\mC)\right] 
\label{eq:expvalbarFrps}
\eeq 
(where  $\tr$ denotes the trace of matrices, to differentiate from the trace of operators $\Tr$), and the squared fidelity deviation is given by 
\begin{align*}
 \left(\Delta\bar{F}\right)^2
=
\frac{1}{4}\Big\{\frac{1}{15} \left[\tr(\mC^2)+\left(\tr(\mC)\right)^2+\tr(\mC\mC^\tp)\right] \nonumber \\
-\left[\frac{1}{3} \tr(\mC)\right]^2\Big\}+\frac{1}{12} \tr(\vv\vv^\tp) .
\end{align*}

Note that different QTPs can result in the same matrix $\mathbb{C}_{\Lambda}$ in~Eq.~\eqref{eq:expvalbarFrps}, producing the same average fidelity.  These protocols in general are not equivalent because they can yield physically distinct output states.
\par

\section{Optimal protocols for fixed average fidelity.}

Let us consider a set of arbitrary teleportation protocols that yield the same average fidelity. 
The following theorem characterizes the optimal protocols within this set:\\
\begin{theorem}
Among all QTPs such that $\expval{\bar{F}}=\alpha \in (0,1]$,
the average trace distance $\expval{\bar{D}_{\text{T}}}$ takes its minimum value for those protocols that align, \ie when the corresponding Bloch vectors of the output states $\hrho^\out_m$ are given by $\vt^{\rm alig}_m=s_m\vt \ \forall\, m$ with $s_m \in (0,1]$ satisfying $\sum_m P_m s_m=2\alpha-1$. These protocols are defined as optimal.
\end{theorem}
{\sl Proof.} For arbitrary QTPs, we have that
\bea
\expval{\bar{D}_{\text{T}}}&=&
\expval{\frac{1}{2}\sum_{m} P_m\,\|\vt-\vt_m\|}\nonumber\\
&\geq&
\expval{\frac{1}{2}\|\vt-\vt_{\Lambda}\|}\geq
1-\expval{F(\hrho^{\inp}, \Lambda(\hrho^{\inp}))}\nonumber\\
&=&\frac{1}{2}
\left(1-\frac{1}{3}\tr(\mC)\right)=1-\expval{\bar{F}},
\label{eq:mainineq}
\eea
where we used consecutively: Jensen's inequality 
$\sum_n P_n\,\|\va_n\|\geq \|\sum_n P_n\,\va_n\|$ with $\sum_n P_n=1$ \ 
(because every norm is a convex function), \ 
$D_{\text{T}}\geq 1-F$~\cite{NielsenChuangBook}, \ $F(\hrho^{\inp}, \Lambda(\hrho^{\inp}))=\bar{F}$ because $\hrho^{\inp}$ is a pure state, and Eq.~\eqref{eq:expvalbarFrps}.
Let us now consider fidelity-equivalent protocols, in the sense that  $\expval{\bar{F}}=\alpha$ is satisfied for  given $\alpha \in (0,1]$. The average trace distance can take different values, with a fixed lower bound,  $\expval{\bar{D}_{\text{T}}}\geq 1-\alpha$, as deduced from Eq.~\eqref{eq:mainineq}. It is straightforward to see that this lower bound is attained by protocols such that $\vt^{\rm alig}_m=s_m\vt \ \forall\, m$ with $\sum_m P_m s_m=2\alpha-1$. \hfill $\square$
\par

Let us now give a comprehensive characterization of the optimal protocols, defined in Theorem~1. In this context, the following result establishes the necessary and sufficient conditions to have a protocol that aligns. 
\par
\noindent 
\begin{theorem}
An arbitrary QTP aligns if and only if it satisfies, for all $m$, that: \ (i)~$\mw_m \vr^{\a}+\vw_m^{\ba}=\vzero$, \ (ii)~$\vr^{\b}= \vzero$ and $\vw_m^{\a}=\vzero$, and  \ (iii)~$\mR_m=s_m\mw_m^{-\tp} \mr^{-\tp}$ with $s_m$ such that $\vt_m=s_m\vt \ \forall\, m$, being $\vt_m$ the Bloch vector of the output state of the protocol.
\end{theorem}
%
\begin{corollary}
The quantum channel associated to a protocol that aligns is characterized by $\mCal=\frac{1}{3}\tr(\mCal)\mid$ and $\vv^{\rm alig}=\vzero$. Thus, this kind of protocols yield null fidelity deviation, $\Delta^{\rm alig}\bar{F}=0$.
\end{corollary}
{\sl Proof. }
The final Bloch vector $\vt_m$ of Bob qubit is given in~\eqref{eq:tm} with $\vt_m^{\b}$ in~\eqref{eq:tmb}.
Therefore, if $\vt_m=s_m\vt $, then $g_m$ 
must be independent of $\vt$ and $\vka_m$ 
must vanish, for all $m$. 
The first condition happens iff the statement~(i) of the theorem is true.
Applying the POVM conditions~\eqref{eq:POVMcona}  and~\eqref{eq:POVMconb} 
to the equations  $\vka_m=\vr^{\b}+\mr^\tp\vw_m^{\a}=\vzero$ $\forall\, m$, we arrive at $\vr^\b=\vzero$, so $\mr^\tp\vw_m^{\a}=0$ $\forall\, m$. 
Because $\vka_m=\vzero$ $\forall\, m$ and $\vr^{\b}=\vzero$, we must have that $\frac{\mR_m\mr^\tp\mw_m^\tp}{g_m(\vt)}=s_m\mid$  in order to align, \ie $\vt=s_m\vt_m$ $\forall\, m$. Therefore, the matrices  $\mr$ and $\mw_m$ must be invertible and from the condition $\mr^\tp\vw_m^{\a}=0$ $\forall\, m$, we get that $\vw_m^{\a}=\vzero$ $\forall\, m$.
 At this point, we demonstrated statement~(ii) of the theorem. 
Note that we have arrived at $g_m(\vt)=1$ $\forall\, m$, that implies $\mR_m\mr^\tp \mw_m^\tp=s_m\mid$. This proves statement~(iii) of the theorem.
\par
Note that statement~(iii) of Theorem~2 implies that $\mCal=\sum_m{P}_m\,s_m\mid=\frac{1}{3}\tr(\mCal)\mid$, 
and $\vka_m=\vzero$ $\forall\, m$ leads to $\vv=\vzero$. These are the statements of Corollary~1. On the other hand, since the lower bound in Eq.~\eqref{eq:mainineq} is achieved for protocols that align,
we have that for average fidelity $\alpha$ it holds, 
\beq
\label{eq:trCalign}
2\alpha -1=\frac{1}{3}\tr(\mCal)=\sum_m\bar{P}_m\,s_m. 
\eeq
\hfill $\square$
\par
Before establishing the next theorem, we recall that under suitable local unitary transformations, \ie
\begin{equation*}
\hrhoc^{ab}=\hat U^{\a}\otimes \hat U^{\b}\hrho^{ab} (\hat U^{\a})^\dagger\otimes (\hat U^{\b})^\dagger ,
\end{equation*}
every two-qubit state $\hrho^{ab}$ can be transformed into a canonical form $\hrhoc^{ab}$, with correlation matrix \ $\mrd=(\mo^\a)\mr(\mo^\b)^\tp=\diag(r_1,r_2,r_3)$, \ being $\mo^\a$ and $\mo^\b$ rotation matrices, and transformed marginal Bloch vectors \ $\vrc^{\a}=\mo^{\a}\vr^{\a}$ and $\vrc^{\b}=\mo^{\b}\vr^{\b}$  \cite{Horodecki1996}.
Furthermore, the positivity condition on the density operators $\hrho^{ab}$ and $\hrhoc^{ab}$, when 
$\vr^{\b}=\vzero$, corresponds to the inequalities  \cite{Gamel2016}:
\bse
\label{eq:faithfulrho2equiv}
\bea
&-2\det(\mrd)-(\|\mrd\|^2-1)\geq \|\vrc^{a}\|^2,\\
&f(r_1,r_2,r_3)\geq
4\|(\vrc^{\a})^\tp\mrd\|^2\nonumber\\
&+\|\vr_{\rm c}^{a}\|^2\,\left[2(1-\|\mrd\|^2)-\|\vr_{\rm c}^{a}\|^2\right],
\label{eq:faithfulrho2cequiv}
\eea
\ese
where $f(r_1,r_2,r_3)=-8\det(\mrd)+(\|\mrd\|^2-1)^2-4\|\mrtd\|^2=(1 - r_1 - r_2 - r_3)(1 - r_1 + r_2 + r_3)(1 + r_1 - r_2 + r_3)(1 + r_1 + r_2 - r_3)$, \ $\|\mrd\|^2=\tr(\mrd^2)$, \ $\mrtd=\det(\mrd)\,\mrd^{-1}$, and \ $-1\leq\det(\mrd)\le 1$ \ (here we are assuming that $\det(\mrd) \not= 0$). Thus, the diagonal elements $r_1,r_2,r_3$ belong to a convex subset, defined by Ineqs.~\eqref{eq:faithfulrho2equiv}, inside the tetrahedron given by inequality~\eqref{eq:faithfulrho2cequiv} with $\vr_{\rm c}^{\a}=\vzero$ \cite{Horodecki1996}.
\par 
We are now in a position to present the following theorem which fully characterizes the QTPs that align:
\par
\noindent 
\begin{theorem}
All QTPs that align verify that: (i)~the POVM states $\homega_m^{\ba\a}$ have correlation
matrices $\mw_m=(\mo_m^{\ba})^\tp\mwmd\mo^{\a}$ with $\mwmd=s_m\mrd^{-1}$ and where  $\mr=(\mo^\a)^\tp\mrd\mo^\b$ is the correlation matrix of the resource state $\hrho^{\a\b}$, with $(\mo^{\a})^\tp$ the rotation matrix 
that simultaneously diagonalizes the positive definite matrices $\mr\mr^\tp$ and $\mw_m^\tp\mw_m$, while  $\mo^{\b}$ and $\mo_m^{\ba}$ are the rotation matrices that diagonalize  $\mr^\tp\mr$ and $\mw_m\mw_m^\tp$ respectively; \ 
(ii)~Bob rotation matrices are of the form $\mR_m=\mo_m^{\ba}\mo^{\b}$ $\forall\, m$;  \ finally, (iii)~the rotation matrices $\mo_m^{\ba}$ must fulfill the  POVM condition (d) that in this case reduces to $\sum \bar{P}_ms_m(\mo_m^{\ba})^\tp=\m0$. 
\end{theorem}
%
\begin{corollary}
All the protocols that align have  $\det(\mrd) < 0$.
\end{corollary}
{\sl Proof. }
From the canonical decomposition of the states $\hrho^{\a\b}$ and $\homega_m^{\ba\a}$ we have that 
$\mr=(\mo^\a)^\tp\mrd\mo^\b$ and $\mw_m=(\mo_m^{\ba})^\tp\mwmd\mo_m^\a$, where 
the columns of $\mo^\a$ are eigenvectors of $\mr\mr^\tp$ and the columns of $\mo_m^{\a}$ are eigenvectors 
of $\mw_m^\tp\mw_m$. Note that $\mr\mr^\tp$ and $\mw_m^\tp\mw_m$ are positive-definite matrices, because 
$\mr$ and $\mw_m$ are full-rank. They are diagonalized by orthogonal matrices. From the orthogonality of $\mR_m$ and condition~(iii) in Theorem~{\it 2},
we get $\mr\mr^\tp\left(\frac{\mw_m}{s_m}\right)^\tp\frac{\mw_m}{s_m}=\mid$, leading to $[\mr\mr^\tp,\mw_m^\tp\mw_m]=\m0$. Then, $\mr\mr^\tp$ and $\mw_m^\tp\mw_m$ are diagonalized 
by a single orthogonal matrix \cite{horn2013}, that we can choose to be one of the possible matrices $(\mo^\a)^\tp$ in the canonical decomposition of $\mr$, \ie $\mr\mr^\tp=(\mo^\a)^\tp\mrd^2\,\mo^\a$ and $\mw_m^\tp\mw_m=(\mo^{\a})^\tp\mwmd^2\mo^{\a}$,
so $\mo_m^{\a}=\mo^\a$ $\forall\, m$. 
Thus, we immediately arrive at $\mwmd^2=s_m^2\mrd^{-2}$.
Finally, we can write $\mR_m=\mo_m^{\ba} \mwmd^{-1}s_m\mrd^{-1}\mo^{b}$, and then
$\mwmd=s_m\mrd^{-1}$ must be true \footnote{The case $\mwmd=-s_m\mrd^{-1}$ is not considered, because $\mR_m$ is a rotation matrix, \ie  $\det(\mR_m)=1$}. This proves statement~(i). Statements~(ii) and~(iii) follow straightforwardly. \hfill $\square$
\par 
From Theorem~3 it is possible to conclude that the only QTP  such that $\vt_m=\vt$ $\forall\, m$ is, up to local unitaries on the qubit systems, the perfect QTP defined by performing a Bell measurement on qubits $\ba$ and $a$ and a Bell state as resource for qubits $a$ and $b$. Specifically, the positivity conditions on the density operators $\hrho^{\a\b}$ and $\homega_m^{\ba\a}$ correspond, respectively, to the set of inequalities \eqref{eq:faithfulrho2equiv} for the matrix elements of $\mrd$  with $\vr_{\rm c}^{\b}=\vzero$, and for the matrix elements of $\mw_{{\rm d}m}=s_m\mrd^{-1}$ with Bloch vectors  
$\vw_{{\rm c}m}^{\ba}=-\mrd^{-1}\vrc^{\a}$ and $\vw_{{\rm c}m}^{\a}=\vzero$ $\forall\, m$ (this follows from conditions~(i) and~(ii) of Theorem~2; see Eq.~\eqref{eq:Gamelmrd2tot} in Appendix~\ref{Apppositivity}). The only solutions to these sets of inequalities, when $s_m=1$ $\forall m$, correspond to $\mrd^{\Bell}=(\mrd^{\Bell})^{-1}=\mw^{\Bell}_{{\rm d}m}\in \{\mr^{\Bell}_{\Phi_+} = -\diag\left(1,1,1\right), \ \mr^{\Bell}_{\Phi_-} = \diag\left(-1,1,1\right), \mr^{\Bell}_{\Psi_+} = \diag\left(1,-1,1\right),   \mr^{\Bell}_{\Psi_-} = \diag\left(1,1,-1\right)\}$ with $\vr_{\rm c}^{\a}=\vzero$ $\forall\, m$. 
These solutions are Bell states for the resource, 
$\hrho_{\rm c}^{\a\b}=\hbeta$, and for the POVM operators, 
$\homega_{{\rm c}m}^{\ba\a}=\hbeta_m$ with $m=1,\ldots,4$,  
in the canonical form. Therefore, from condition~(i) of Theorem~3 we have that the 
correlation matrix of $\homega_{{\rm c}m}^{\ba\a}$ is \  $\mw_m=\mw_{{\rm c}m}=(\mo_m^{\ba})^\tp\mwmd\mo^{\a}$
with $\mo^{\a}=\mid$ and
$\mo_m^{\ba}=\mb_{m}^{\ba}\in$ $\{\diag\left(1,1,1\right)$, $\diag\left(1,-1,-1\right)$, $\diag\left(-1,1,-1\right)$, $\diag\left(-1,-1,1\right)\}$, that are the only  diagonal orthogonal matrices in $\Rset^{3\times 3}$
with $\det(\mb_{m}^{\ba})=1$. Note that these matrices satisfy condition~(iii) of Theorem~{\it 3}.
All perfect QTPs, therefore, are those with resource state $\hrho^{\a\b}=\hat U^{\a}\otimes\hat U^{\b}\,\beta \,(\hat U^{\a})^\dagger\otimes(\hat U^{\b})^\dagger$, being $\mr=(\mo^{\a})^\tp\mrd^{\Bell}\mo^{\b}$ its correlation matrix, and with a POVM composed by $\homega^{\ba\a}_m=\hat U^{\ba}\otimes \hat U^ { \a}\homega^{\ba\a}_{{\rm c}m} (\hat U^{\ba})^\dagger\otimes (\hat U^{\a})^\dagger$, with $\homega_{{\rm c}m}^{\ba\a}=\hbeta_m$, whose correlation matrices are
$\mw=(\mo^{\ba})^\tp\mb^{\ba}_{m}\mrd^{\Bell}\mo^{\a}$. 
\par
It is worth noting that, according to Theorem~3, for teleportation protocols that align, the POVM states can be written as $\homega_m^{\ba\a}=\hat U_m^{\ba}\otimes \hat U^ { \a}\homega^{\ba\a}_{{\rm c}m} (\hat U_m^{\ba})^\dagger\otimes (\hat U^{\a})^\dagger$, where $\hat U^ { \a}$ is one of the local unitary operations that carries $\hrho^{\a\b}$ into its canonical form. Therefore, the Bob qubit state $\hrho_m^{\b}$, after Alice measurement, does not depend on $\hat U^{\a}$. So, we can ignore this local unitary operation.

Now, let us examine the scenario where $\hat U_m^{\ba}$ is a unitary matrix such that 
$\hat{U}^{\ba}_m\, \vn^\tp\hbsigma \, (\hat{U}^{\ba}_m)^\dagger = 
\left(\mo^{\ba}_m \vn \right)^\tp \hbsigma$ with $\mo^{\ba}_m$ a diagonal matrix.
In the case of diagonal matrices $\mo_m^{\ba}$, the only possible way to satisfy condition~(iii) of Theorem~ {\it 3} is when
$\mo_m^{\ba}=\mb_{m}^{\ba}$ and $s_m=s$ for $m=1,\ldots,4$. 
 
Therefore, for these particular protocols considered, the resource state has a correlation matrix $\mr=\mrd\mo^b$ and the POVM states have
$\mw_m=\mw_{{\rm c}m}=\mb_{m}^{\ba} s \, \mrd^{-1}$ with $m=1,\ldots,4$, \ie  $\homega^{\ba\a}_{m}=\homega^{\ba\a}_{{\rm c}m}$. 
We refer to this kind of protocols as deterministic quantum teleportation protocols (DQTPs) that align. 
For these protocols the Bloch vectors of the reduced states $\hrho^{\a}$ and $\homega^{\ba}_m$ are, respectively, $\vr^{\a}=\vrc^{\a}$ and $\vw_m^{\ba}=(\mb_m^{\ba})^\tp\vw_{{\rm c}m}^{\ba}$
with $\vw_{{\rm c}m}^{\ba}=\vw_{{\rm c}}^{\ba}=-s\,\mrd^{-1}\vrc^{\a}$ for $m=1,\ldots,4$. 
\par
The perfect QTP, $s=1$, is a special case of a DQTP that aligns corresponding to $\vrc^{\a}=\vzero$ and $\mrd=\mrd^{\Bell}$.

In the case of imperfect alignment of the DQTP, where $s<1$, the set of allowed values for the diagonal elements of $\mrd$ and $s\,\mrd^{-1}$, as determined by the positivity conditions for the density operators of the resource and POVM states, is quite extensive. 

Let us consider, as an example, all the protocols that align for different values of $s$ and Bloch vectors $\vr_{\rm c}^{\a}$ in the scenario $s_m =s$ for all  $m$. Figure 1 illustrates the sets of values $(r_1, r_2, r_3)$, represented by the shaded red volume, for which there exists a POVM that aligns for different values of $\|\vr_{\rm c}^{\a}\|$ and considering two different values of $s$.  These regions are determined by the positivity conditions on $\hrho^{\a\b}$ and $\homega_m^{\ba\a}$ (see inequalities \eqref{eq:faithfulrho2equiv}). It can be observed that, as $\|\vr_{\rm c}^{\a}\|$ increases, the set of solutions becomes smaller. In panels (a) to (d) we consider $s=0.8$; notice that in panel (d), corresponding to $\|\vr_{\rm c}^{\a}\|=0.2$, there is no solution. However, when we reduce the average fidelity value as in panels (e) to (h) where we take $s=0.7$, a tiny set of solutions appears for $\|\vr_{\rm c}^{\a}\|=0.2$ (indicated by solid arrows (red online) in panel (h)).

\begin{figure*}
   \centering
  \includegraphics[width=16cm,height=8cm]{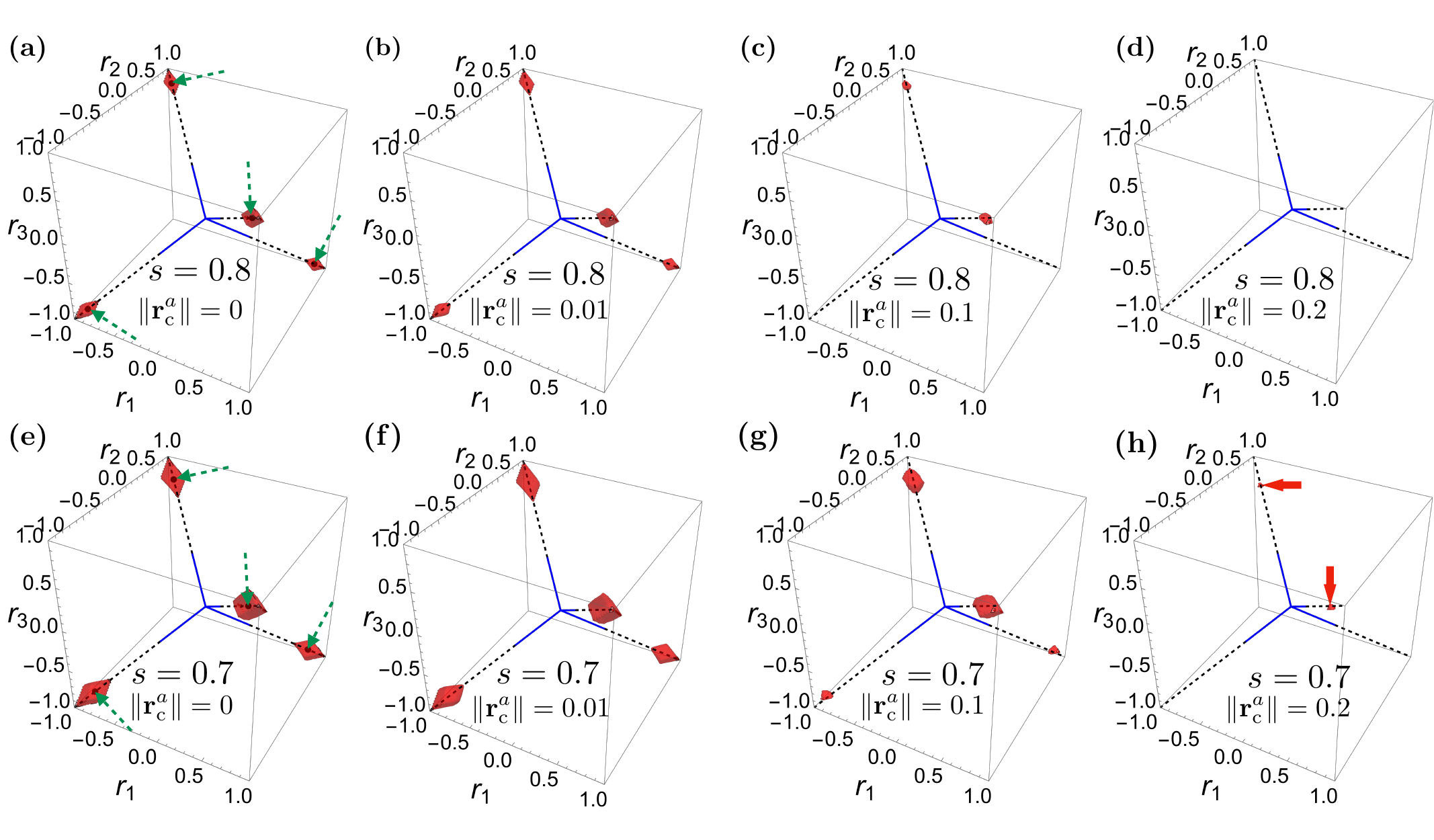}
   \caption{Diagonal matrix elements $(r_1,r_2,r_3)$ of DQTPs that align, with fixed value of average fidelity 
  $\expval{\bar{F}}^{\rm alig}=\frac{1}{2}(1+s)=\alpha$, for different values of $s$ and Bloch vectors $\vr_{\rm c}^{\a}$ (see text for explanation).  The angular spherical coordinates of $\vr_{\rm c}^{\a}$ are $\theta=\phi=0$ in panels (a)--(d), and $\theta=\phi=\frac{\pi}{2}$ in panels (e)--(h). In (d) there is no solution; in (h) the solid arrows (red online) indicate the tiny set of solutions. The lines (blue online) correspond to the four types of Werner states $\hat W=\frac{(1-p)}{4}\hid \otimes\hid+p\,\hbeta$ where 
 $\hbeta$ is one of the four Bell states. 
The solid lines correspond to separable states, \ie $-\frac{1}{3}\leq p\leq \frac{1}{3}$, and the dashed  lines to entangled states, \ie  $\frac{1}{3}< p\leq 1$. 
In  (a) and (e), the dashed arrows (green online) indicate the special cases with $p=s^{\frac{1}{2}}$.}
\label{fig1}
\end{figure*}
\par 

\section{Noise in DQTP that align}

For a deterministic QTP meeting the conditions in Theorems~$2$ and~$3$, from Eq.~\eqref{eq:trCalign} we have that 
$\expval{\bar{F}}^{\rm alig}=\frac{1}{2}(1+s)=\alpha$. Therefore, we see that for a fixed value $\alpha<1$, \ie fixed $s<1$, there exist different DQTPs that align giving rise to the same average fidelity (see Fig.~\ref{fig1}).
These different protocols can be identified as the action of one-qubit channel over 
the perfect DQTP that aligns:
$\hrho_{\rm c}^{\a\b}=(\varepsilon^{\a}\otimes \varepsilon^{\b})[\hbeta]$ and $\homega^{\ba\a}_{{\rm c}m}=(\varepsilon^{\ba}\otimes \varepsilon^{\a})[\hbeta_m]$ with $m=1,\ldots,4$. 

\par
A generic one-qubit channel $\varepsilon$ can be described by the affine transformation $\vt^{\rm out}=\mathbb{\Lambda}\vt^{\rm in}+{\bf v}$ of the vectors $\vt^{\rm in}$ in the Bloch sphere, where $\mathbb{\Lambda}$ and ${\bf v}$ are the matrix and the translation vector of the channel respectively \cite{Zyczkowski-book}.
\par
Using the result in Appendix~\ref{Appnoise} it is shown that the correlation matrix of $\hrho_{\rm c}^{\a\b}=(\varepsilon^{\a}\otimes \varepsilon^{\b})[\hbeta]$
is $\mrd=\mLambda_d^\a\;\mLambda_d^\b\;\mrd^\Bell$, where $\mathbb{\Lambda}_{\rm d}^{\a}$ and $\mathbb{\Lambda}_{\rm d}^{\b}$ are the diagonal matrices of the affine description of the channels $\varepsilon^{\a}$ and $\varepsilon^{\b}$, respectively.
Note that, because the values of the diagonal entries of $\mathbb{\Lambda}_{\rm d}^{\a}\mathbb{\Lambda}_{\rm d}^{\b}$ are inside the tetrahedron of allowed values for channels, the values of the diagonal entries of $\mrd=\mLambda_d^\a\;\mLambda_d^\b\;\mrd^\Bell$ are inside the tetrahedron of allowed values for correlation matrices of two-qubit states~\cite{Zyczkowski-book}.
The Bloch vectors of the reduced states of $\hrho_{\rm c}^{\a\b}=(\varepsilon^{\a}\otimes \varepsilon^{\b})[\hbeta]$ are 
$\vvv^{\a}=\vrc^{\a}$ and $\vvv^{\b}=\vrc^{\b}=\vzero$, with $\vvv^{\a}$ and $\vvv^{\b}$ the affine vectors of the channels 
$\varepsilon^{\a}$ and $\varepsilon^{\b}$ respectively. It follows that the channel $\varepsilon^{\b}$ must be unital.
The correlation matrix of the POVM states $\homega^{\ba\a}_{{\rm c}m}=(\varepsilon^{\ba}\otimes \varepsilon^{\a})[\hbeta_m]$
are $\mw_{{\rm c}m}=s\,\mb_m^{\ba}\;\mrd^{-1}=
(\mLambda_d^\a\,\mLambda_d^\b)^{-1}\,s\;\mb_m^{\ba}\;\mrd^\Bell =\mLambda_d^{\ba}\,\mLambda_d^{\a}\mb_m^{\ba}\;\mrd^\Bell $, with $\mathbb{\Lambda}_{\rm d}^{\ba}$ and $\mathbb{\Lambda}_{\rm d}^{\a}$ the matrices of the affine description of 
$\varepsilon^{\ba}$ and $\varepsilon^{\a}$ respectively. Then, we arrive at first  condition 
\beq
\label{condition1}
\mLambda_{\rm d}^{\ba}  (\mLambda_{\rm d}^\a)^2\mLambda_d^\b=
s\;\mid. 
\eeq
The second condition, correlating channels on qubits $\ba$, $\a$ and $\b$, is  
\beq
\label{condition2} 
{\bf v}^{\ba}=\vw_{{\rm c}}^{\ba}=-s\;(\mLambda_d^\a\;\mLambda_d^\b)^{-1}\;\mrd^\Bell \;\vrc^{\a}.
\eeq
Notice that this last condition disappears if the channel $\varepsilon^{\a}$ is unital, \ie with affine vector ${\bf v}^{\a}=\vrc^{\a}=\vzero$. Thus in this case all the three qubit channels $\varepsilon^{\a}$, $\varepsilon^{\ba}$ and $\varepsilon^{\b}$ must be unital in order to have a DQTP that aligns. It is worth noting that all more common noisy one qubit quantum channels are of this kind \cite{Zyczkowski-book}. 
\par
Conditions~\eqref{condition1} and~\eqref{condition2} show that in general the channels $\varepsilon^{\ba}$, $\varepsilon^{\a}$ and $\varepsilon^{\b}$
are correlated. Uncorrelated solutions of~\eqref{condition1} occur only when the channel matrices are independent. 
If none of the channels is the identity (no noise), uncorrelated solutions are only achieved when all the channels are the same, \ie $\mLambda_{\rm d}^{\ba} =\mLambda_d^\b=\mLambda_d^\a=\mLambda_{\rm d}$, and $\mLambda_{\rm d}=s^{\frac{1}{4}}\mid$ (which, in turns, defines a depolarizing channel \cite{Zyczkowski-book}). Because these channels are unital, $\vr_{\rm c}^{\a}=\vzero$ so $\vw_{{\rm c}}^{\ba}=\vzero$, condition~~\eqref{condition2} is automatically satisfied.
In this case, both the resource and POVM states are Werner states, \ie $\hrho^{\a\b}=\hat W$ and $\homega^{\ba\a}_m=\hat W_m$ where $\hat W_{m}=\frac{(1-p)}{4}\hid \otimes\hid+p\,\hat\beta_m$, with $m=1,\ldots,4$, and $\hat W$ being one the previous states. The noise parameter $p$ satisfies: $p=s^{\frac{1}{2}}$, for $\hat W$ and $\hat W_m$ $\forall\, m$.  
For each fixed value of $s$, \ie fixed average fidelity, these DQTPs that align
are spotted in panels $(a)$ and $(e)$ of Fig.~\ref{fig1} with dashed (green online) arrows.
These are also the solutions of DQTPs that align corresponding to uncorrelated channels but with noise only in one or two of the qubit systems of the protocol. In this case, the only difference is that the depolarizing channels have a matrix $\mLambda_{\rm d}=s^{\frac{1}{2}}\mid$.
\par
It is worth noting that DQTP that align corresponding to uncorrelated noise in all the qubits are formed by entangled Werner states when $\frac{1}{3}\leq p={s}^{\frac{1}{2}}\leq 1$, and by separable when $0 < p={s}^{\frac{1}{2}}\leq \frac{1}{3}$. In the case of separable states, the average fidelity of the protocols ranges  $0<\expval{\bar{F}}^{\rm alig}\leq \frac{5}{9}< \frac{2}{3}=\expval{\bar{F}}^{\rm cl}$, with $\expval{\bar{F}}^{\rm cl}$ the average fidelity corresponding to the classical protocol \cite{MassarPopescu1995,Vidal1999}.  These show that entanglement is needed to surpass the average fidelity of the classical protocol, both in the resource state and also in the POVM states.
\par
DQTPs that align with Werner states, \ie 
$\hrho^{\a\b}=\hat W$ and $\homega^{\ba\a}_m=\hat W_m$, exist if 
the parameter that define all $\hat W_m$ states is $p^\prime=\frac{s}{p}$, where $p$ is the parameter that define $\hat W$.
In these protocols, the correlation matrix of $\hrho^{\a\b}=\hat W$ is $\mrd=p \mrd^{\Bell}$, and those of
$\homega^{\ba\a}_m=\hat W_m$  are $\mw_{{\rm c}m}=\mb_{m}^{\ba}\frac{s}{p}\mrd^{\Bell}$. 
Replacing in the positivity  condition~\eqref{eq:faithfulrho2cequiv}, $\mrd$  by $\mw_{{\rm d}m}=\frac{s}{p}\mrd^{\Bell}$  and $\vr^{\a}_{\rm c}$ by $\vw_{{\rm c}m}^{\ba}=-\mrd^{-1}\vrc^{\a}=\vzero$ , we can rewrite this inequality as $p^8\,(p-s)^3\,(p+3s) \geq 0$. 
The solution of this inequality, together with $0<s\leq 1$ and $-\frac{1}{3}\leq p \leq 1$~\footnote{These two necessary inequalities guarantee that the matrix elements of  $\mrd=p \mrd^{\Bell}$ satisfy the positivity condition in~\eqref{eq:faithfulrho2cequiv} with $\vr^{\a}_{\rm c}=\vzero$. }, corresponds to two cases:
case~{\it (I)} when $s\leq p\leq 1$ with $0< s\leq 1$, and case~{\it (II)} when $-\frac{1}{3}\leq p\leq -3 s$ with 
$0<s\leq \frac{1}{9}$.
We stress that only when $p^\prime=p={s}^{\frac{1}{2}}$ the DQTP that align is associated with uncorrelated noise in the qubits.
This is a particular solution included in the case~{\it (I)}. 
Also, note that the DQTPs that align with Werner states become standard noisy QTPs when $p=s$ and $s<1$ so $\hat W_m=\hat \beta_m$ $\forall\, m$. When $s=1$ it becomes the perfect QTP.
All these DQTPs that align with Werner states need entanglement, both in the resource state and also in the POVM states, to surpass the average fidelity of the classical protocol.
\par

\section{Conclusions.}
  
We demonstrate that the optimal quantum teleportation protocols over pure random states, with a fixed average fidelity, are those that align the Bloch vectors of the input and output states. In other words, $\vt_m=s_m\vt$, where $s_m$ is independent of the initial Bloch vector $\vt$, for any outcome $m$ of Alice measurement. This alignment results in output states that are diagonal in the same basis as the initial state. Besides, these protocols effectively act as depolarizing channels, $\hrho_m=\Lambda^{\rm dep}_m(\hrho^{\rm in})$, for each $m$.  We characterize all the resource states and POVM measures of these optimal protocols, which in turn determine the rotation operation in the output state of the protocols.
\par
A remarkable type of aligned QTP is when $s_m=s$ for all $m$. These deterministic protocols are particularly relevant as they emerge when attempting to implement the perfect QTP under the influence of correlated noise in qubit systems. Among these protocols, we demonstrate the existence of one with uncorrelated noise, corresponding to the same depolarizing channel in the qubits. The amount of noise in this protocol determines the average fidelity of the teleportation process, a situation commonly encountered in experimental implementations \cite{Cai2021}. Therefore, in such experimental scenarios, we establish that the optimal QTP involves preparing a Bell state as the resource state and employing a Bell measurement as a POVM.

\hfill

%
\acknowledgments{ FT acknowledges financial support from the Brazilian agency INCT-Informa\c{c}\~ao Qu\^antica. 
DGB, GMB, APM, and MP acknowledge Consejo Nacional de Investigaciones Científicas y Técnicas (CONICET), Argentina, for financial support. 
DGB, GMB, and MP are also grateful to
Universidad Nacional de La Plata (UNLP), Argentina. APM acknowledges partial support from SeCyT, Universidad Nacional de Córdoba (UNC), Argentina.
}

\appendix

\section{Positivity conditions on the density operators $\hrho^{\a\b}$ and 
$\homega_m^{\ba\a}$ that satisfy Theorem~2} 
\label{Apppositivity}

Here we explicitly write down the inequalities that define the positivity conditions on the density operators $\hrho^{\a\b}$ and 
$\homega_m^{\ba\a}$, that satisfy Theorem~2, following Ref. \cite{Gamel2016}.

The positivity conditions on density operators $\hrho^{\a\b}$ of the form in~\eqref{eq:Fano2} were given in \cite{Gamel2016}.
When the marginal Bloch vector $\vr^{\b}$ is null these inequalities are:
\begin{widetext}
\bse
\bea
3-\|\mrd\|^2&\geq& \|\vr_{\rm c}^{\a}\|^2\\
-2\det(\mrd)-(\|\mrd\|^2-1)&\geq& \|\vr^{a}_{\rm c}\|^2
\label{eq:Gamelrmd3a}\\
-8\det(\mrd)+(\|\mrd\|^2-1)^2-4\|\mrtd\|^2&\geq&
4\|\mrd\vr_{\rm c}^{\a}\|^2+\|\vr^{a}_{\rm c}\|^2 [2(1-\|\mrd\|^2)-\|\vr^{a}_{\rm c}\|^2],
\label{eq:Gamelrmd3}
\eea
\ese
\end{widetext}
where $\mrtd=\det(\mrd)\,\mrd^{-1}$.  The correlation matrix $\mrd=(\mo^\a)\mr(\mo^\b)^\tp=\diag(r_1,r_2,r_3)$ and the marginal Bloch vector $\vrc^{\a}=\mo^{\a}\vr^{\a}$  
correspond to 
the state in the canonical form $\hrhoc^{\a\b}$.  It is straightforward to show that when the matrix $\mrd$ is invertible, \ie $\det(\mrd)\neq 0$, 
the first equation is redundant.

Equivalently, the relevant positivity conditions on the density operators $\homega_m^{\ba\a}$ that satisfy Theorem~2 are:
\begin{widetext}
\bse
\label{eq:Gamelforwm}
\bea
-2\det(\mwmd)-(\|\mwmd\|^2-1)&\geq& \|\vw_{m,\rm c}^{\ba}\|^2\\
-8\det(\mwmd)+(\|\mwmd\|^2-1)^2-4\|\mwtmd\|^2&\geq&
4\|\mwmd\vw_{m,\rm c}^{\ba}\|^2+
\|\vw_{m,\rm c}^{\ba}\|^2[2(1-\|\mwmd\|^2)-\|\vw_{m,\rm c}^{\ba}\|^2].
\eea
\ese
\end{widetext}
Now we know that 

\begin{equation*}
\mwmd=s_m\mrd^{-1} ,
\end{equation*}
\begin{equation*}  
\mwtmd=\det(\mwmd)\,\mwmd^{-1}=\frac{s_m^3}{\det(\mrd)}\frac{1}{s_m}\mrd=\frac{s_m^2}{\det(\mrd)}\mrd , 
\end{equation*} 
and 
\begin{equation*}
\vw_{m,\rm c}^{\ba}=-s_m\mrd^{-1}\vr_{\rm c}^{\a} .
\end{equation*}
Therefore 

\begin{equation*}
\|\mwmd\|^2=s_m^2\|\mrd^{-1}\|^2=\frac{s_m^2}{(\det(\mrd))^2}\|\mrtd\|^2 , 
\end{equation*}
\begin{equation*}
\|\mwtmd\|^2=\frac{s_m^4}{(\det(\mrd))^2}\|\mrd\|^2 , 
\end{equation*}
\begin{equation*}
\|\vw_{m,\rm c}^{\ba}\|^2=s_m^2\|\mrd^{-1}\vr_{\rm c}^{\a}\|^2=\frac{s_m^2}{(\det(\mrd))^2}\|\mrtd\vr_{\rm c}^{\a}\|^2 , 
\end{equation*} 
and 
\begin{equation*}
\|\mwmd\,\vw_{m,\rm c}^{\ba}\|^2=s_m^4\|\mrd^{-2}\vr_{\rm c}^{\a}\|^2=
\frac{s_m^4}{(\det(\mrd))^4}\|\mrtd^{2}\vr_{\rm c}^{\a}\|^2 .
\end{equation*}
Replacing these expressions into~Eq.\eqref{eq:Gamelforwm} we arrive at the set of inequalities:

\begin{widetext}
\bse
\label{eq:positivitywaab}
\bea
-2s_m^3\det(\mrd)-\left(s_m^2\|\mrtd\|^2-(\det(\mrd))^2\right)&\geq&
s_m^2\|\mrtd\vr_{\rm c}^{\a}\|^2,\\
-8s_m^3(\det(\mrd))^3+\left(s_m^2\|\mrtd\|^2-(\det(\mrd))^2\right)^2-
4s_m^4(\det(\mrd))^2\|\mrd\|^2
&\geq&\nonumber\\ 4s_m^4\|\mrtd^{2}\vr_{\rm c}^{\a}\|^2-
s_m^2\|\mrtd\vr_{\rm c}^{\a}\|^2
\left(2\left(s_m^2\|\mrtd\|^2-(\det(\mrd))^2\right)+s_m^2\|\mrtd\vr_{\rm c}^{\a}\|^2\right).
\eea
\ese
\end{widetext}
Therefore, for given values of the Bloch vector $\vr_{\rm c}^{\a}$ and the parameter $s_m$, the set of allowed values of the matrix elements $r_i$ with $i=1,2,3$ are defined by the inequalities~\eqref{eq:Gamelrmd3a},~\eqref{eq:Gamelrmd3} and~\eqref{eq:positivitywaab}, \ie 
\begin{widetext}
\bse
\label{eq:Gamelmrd2tot}
\bea
-2\det(\mrd)-(\|\mrd\|^2-1)&\geq& \|\vr^{a}_{\rm c}\|^2
\label{eq:Gamelmrd2}\\
-8\det(\mrd)+(\|\mrd\|^2-1)^2-4\|\mrtd\|^2&\geq&\nonumber\\
4\|\mrd\vr_{\rm c}^{\a}\|^2+\|\vr^{a}_{\rm c}\|^2(2(1-\|\mrd\|^2)-\|\vr^{a}_{\rm c}\|^2)\label{eq:Gamelmrd3}\\
-2s_m^3\det(\mrd)-\left(s_m^2\|\mrtd\|^2-(\det(\mrd))^2\right)&\geq&
s_m^2\|\mrtd\vr_{\rm c}^{\a}\|^2\label{eq:Gamelmwd2}\\
-8s_m^3(\det(\mrd))^3+\left(s_m^2\|\mrtd\|^2-(\det(\mrd))^2\right)^2-
4s_m^4(\det(\mrd))^2\|\mrd\|^2&\geq&\nonumber\\
\label{eq:Gamelmwd3}\geq 4s_m^4\|\mrtd^{2}\vr_{\rm c}^{\a}\|^2-
s_m^2\|\mrtd\vr_{\rm c}^{\a}\|^2
\left(2\left(s_m^2\|\mrtd\|^2-(\det(\mrd))^2\right)+s_m^2\|\mrtd\vr_{\rm c}^{\a}\|^2\right)&&.
\eea
\ese
\end{widetext}
\par
Note that the  l.h.s. of inequality~\eqref{eq:Gamelmrd3} is 
\begin{widetext}
\beq
f(r_1,r_2,r_3)=(1 - r_1 - r_2 - r_3)(1 - r_1 + r_2 + r_3)(1 + r_1 - r_2 + r_3)(1 + r_1 + r_2 - r_3)=
-8\det(\mrd)+(\|\mrd\|^2-1)^2-4\|\mrtd\|^2.
\eeq
\end{widetext}

\section{Calculation of the Fano form of $(\varepsilon^\a\otimes\varepsilon^\b)[\hrho^{ab}]$ }
\label{Appnoise}

Here we show the action of  local arbitrary one-qubit channels on a two-qubit state given in the Fano form~\eqref{eq:Fano2}. An analogous calculation with only one{\color{red}-}qubit channel was performed in  Ref.~\cite{Shahbeigi2018}.
 \begin{lemma}
 Let $\varepsilon^\a$ and $\varepsilon^\b$ be  one-qubit channels described by the affine parameters $\Lambda^\a$, ${\bf v}^\a$,
 and $\Lambda^\b$, ${\bf v}^\b$, respectively, and let  $\hrho^{ab}$ be an arbitrary two-qubit state given in Fano form in Eq.~\eqref{eq:Fano2}, then 
 \begin{widetext}
  \bea
(\varepsilon^\a\otimes\varepsilon^\b)[\hrho^{ab}]\! 
&=&
 \!\frac{1}{4}\!\left(\hid\otimes\hid+((\vr^{\a})^\tp\mLambda^\a+({\bf v}^\a)^\tp)\hbsigma\otimes \hid
 +\hid\otimes ((\vr^{\b})^\tp\mLambda^\b+({\bf v}^\b)^\tp)\hbsigma\right.\nonumber\\
 && 
+\left. \sum_{i=1}^3\sum_{j=1}^3 
 \left( {\bf v}^\a  ({\bf v}^\b)^\tp +
 (\mLambda^\a)^\tp\vr^\a  ({\bf v}^\b)^\tp +{\bf v}^\a_i (\vr^\b)^\tp\mLambda^\b+
 (\mLambda^\a)^\tp \;\mr\; \mLambda^\b\right)_{ij}\hsigma_i\otimes\hsigma_j
\right).
\eea
\end{widetext}
 \end{lemma} 
Using the linear property of the quantum channels, we get 
\begin{widetext}
\bea
(\varepsilon^\a\otimes\varepsilon^\b)[\hrho^{ab}]\!&=&\!\frac{1}{4}\!\left(\varepsilon^\a[\hid]\otimes\varepsilon^\b[\hid] +
\varepsilon^\a[(\vr^{\a})^\tp \hbsigma] \otimes \varepsilon^\b[\hid]\!+
\varepsilon^\a[\hid] \otimes  \varepsilon^\b[(\vr^{b})^\tp \hbsigma]+
\!\sum_{i,j=1}^3\!\mr_{ij}\;
 \varepsilon^\b[\hsigma_{i}]\otimes \varepsilon^\b[\hsigma_{j}]\right)\nonumber\\
 &=&
 \!\frac{1}{4}\!\left(\hid\otimes\hid+((\vr^{\a})^\tp\mLambda^\a+({\bf v}^\a)^\tp)\hbsigma\otimes \hid
 +\hid\otimes ((\vr^{\b})^\tp\mLambda^\b+({\bf v}^\b)^\tp)\hbsigma\right.\nonumber\\
 && ({\bf v}^\a)^\tp\hbsigma\otimes ({\bf v}^\b)^\tp\hbsigma+ 
  (\vr^\a)^\tp\mLambda^\a\hbsigma\otimes ({\bf v}^\b)^\tp\hbsigma+
   ({\bf v}^\a)^\tp\hbsigma \otimes (\vr^\b)^\tp\mLambda^\b\hbsigma \nonumber\\
 &&\left.+
 \sum_{k,l=1}^3
 \left((\mLambda^\a)^\tp \;\mr\; \mLambda^\b\right)_{kl}   \hsigma_{k}\otimes \hsigma_{l}\right)\nonumber\\
 &=&
 \!\frac{1}{4}\!\left(\hid\otimes\hid+((\vr^{\a})^\tp\mLambda^\a+({\bf v}^\a)^\tp)\hbsigma\otimes \hid
 +\hid\otimes ((\vr^{\b})^\tp\mLambda^\b+({\bf v}^\b)^\tp)\hbsigma\right.\nonumber\\
 && 
+\left. \sum_{i=1}^3\sum_{j=1}^3 
 \left( {\bf v}^\a  ({\bf v}^\b)^\tp +
 (\mLambda^\a)^\tp\vr^\a  ({\bf v}^\b)^\tp +{\bf v}^\a_i (\vr^\b)^\tp\mLambda^\b+
 (\mLambda^\a)^\tp \;\mr\; \mLambda^\b\right)_{ij}\hsigma_i\otimes\hsigma_j
\right)
\label{doschaneles}
\eea
\end{widetext}
where we used that \ $\varepsilon[\hid]=\hid+({\bf v})^\tp\hbsigma$ \ and \ $\varepsilon[\hsigma_i]=\sum_{j=1}^3\mLambda_{ij}\hsigma_j$ \ 
(so $\varepsilon[\hbsigma]=\mLambda\hbsigma$) that can be easily proven using the affine representation of $\varepsilon$.




%

\end{document}